\documentclass[12pt,onecolumn]{article}

\usepackage{multicol}
\usepackage{graphicx}
\usepackage{amsmath}
\usepackage{lipsum}
\usepackage{color}
\usepackage{siunitx}
\usepackage{authblk}
\usepackage{subfigure}
\usepackage[margin=2cm]{geometry}

\title{\textbf{Damping of liquid sloshing by foams: from everyday observations to liquid transport}}

\author[1,2]{Jean Cappello}
\author[1,3,\footnote{alban.sauret@saint-gobain.com}]{Alban Sauret}
\author[1]{Fran\c{c}ois Boulogne}
\author[4]{Emilie Dressaire}
\author[1,\footnote{hastone@princeton.edu}]{Howard A. Stone}
\affil[1]{\small{Department of Mechanical and Aerospace Engineering, Princeton University, Princeton, New Jersey 08544, USA}}
\affil[2]{Ecole Normale Sup\'erieure de Cachan, 94235 Cachan, France}
\affil[3]{Surface du Verre et Interfaces, UMR 125, 93303 Aubervilliers, France}
\affil[4]{Department of Mechanical and Aerospace Engineering, New York University Polytechnic School of Engineering, Brooklyn, New York 11201, USA}

\small{\date{\today}}

\begin{document}

        \maketitle
        \begin{abstract}
           We perform experiments on the sloshing dynamics of liquids in a rectangular container submitted to an impulse. We show that when foam is placed on top of the liquid the oscillations of the free interface are significantly damped. The ability to reduce sloshing and associated splashing could find applications in numerous industrial processes involving liquid transport.
        \end{abstract}

\bigskip
\bigskip
\bigskip
When a container is set in motion, the free surface of the liquid starts to slosh, i.e. oscillate. For a frequency of motion corresponding to the resonant frequency of the surface wave, the amplitude of the waves can increase significantly and if the amplitude of sloshing is large enough splashing and/or drop formation are possible \cite{ibrahim,faltinsen}. The sloshing dynamics lead to challenging technical constraints in various applications. For example, sloshing leads to considerable pressure forces on the walls of the containers used for transport of oil and liquefied gas \cite{kim}. Therefore, the characterization of the amplitude of the generated waves and the investigation of methods to damp sloshing such as solid foams \cite{patent} or baffles \cite{baffles} are important. 

\begin{figure}[h!]
  \begin{center}
      \subfigure[]{\includegraphics[height=5.5cm]{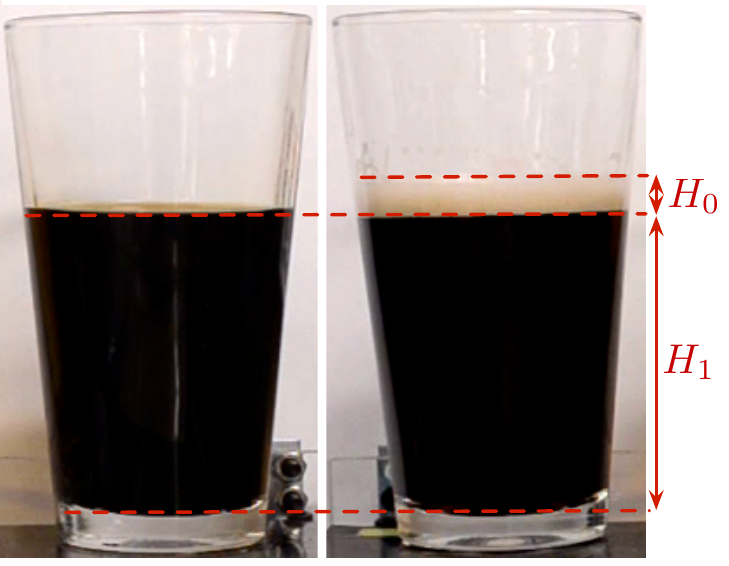}} \qquad \qquad
       \subfigure[]{\includegraphics[height=5.5cm]{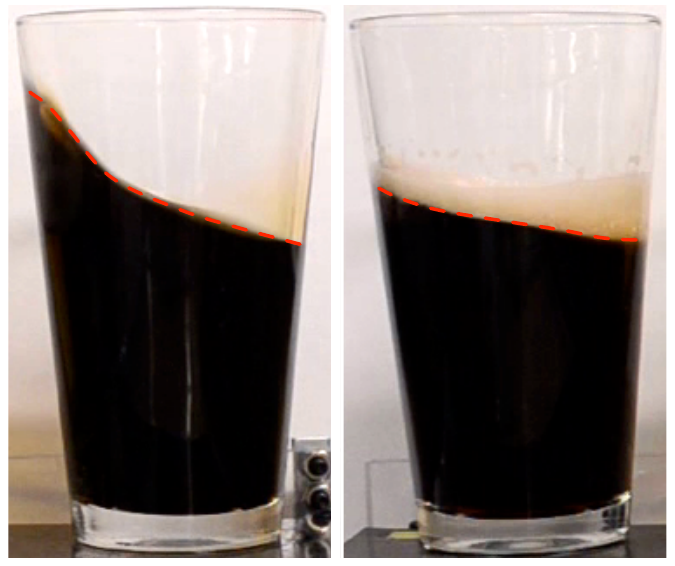}}
     \centering\caption{Photographs of a glass of coffee (left) and a glass of beer (right). The glass is filled to a height $H_1$ of liquid and a foam layer of thickness $H_0$ lies in the top of the beer. (a) At rest, the free interface remains horizontal and (b) after an impulse motion the liquid starts to oscillate. The same impulse is applied to both systems, here shown after the first oscillations.}
     \label{fig:1}
   \end{center}
\end{figure}

Such effects are also observed when a glass of water is handled carelessly and the fluid sloshes or even spills over the rims of the container \cite{mayer}. It is a common observation that beer does not slosh as readily as water, which suggests that the presence of foam could be used to damp sloshing, as illustrated in Figure \ref{fig:1}. In this communication, we consider the effect on sloshing of a liquid foam placed on top of liquid in a rectangular cell.  We present here experimental visualizations of the motion of the free interface and its time evolution.

\section{Experiments}

The experimental setup consists of a rectangular cell of height $H=92\,\rm{mm}$, length $L=70\,\rm{mm}$ and width \mbox{$w=16$ mm}. The vertical and bottom walls are made of glass plates and a rigid rubber sheet, respectively. The cell, the camera and the LED Panel for imaging are set on a moving stage. A mechanical vibrator (LDS 319024-3) controlled by a function generator (Stanford research system Model DS345) and an amplifier (LDS PA100E) provide the impulse. The motion of the interface is recorded using a high-speed camera (Phantom V9.1) at 100 frames per second.

We consider the sloshing dynamics with and without foam using a solution made of $90$ vol \% water, $5$ vol \% glycerol and $5$ vol \% of a commercial surfactant (Dawn dish-washing liquid). The viscosity is approximately $(1.4 \pm 0.1)\times 10^{-6} \,{\rm m^2 \,\, s}$. This mixture forms a stable foam with a low surface modulus \cite{foam_book,denkov}. We use a syringe pump at flow rate $Q = 20\, \mu\rm{L\,\,min^{-1}}$ with a $2.5$ mm diameter needle inserted in the rubber sheet to generate a foam of monodisperse bubbles of  diameter $D=3$ mm. The experiments are performed on short time scales, typically less than few minutes, to avoid aging of the foam.

\section{Results and perspectives}

\begin{figure}[h!]  \begin{center}
\includegraphics[width=19cm]{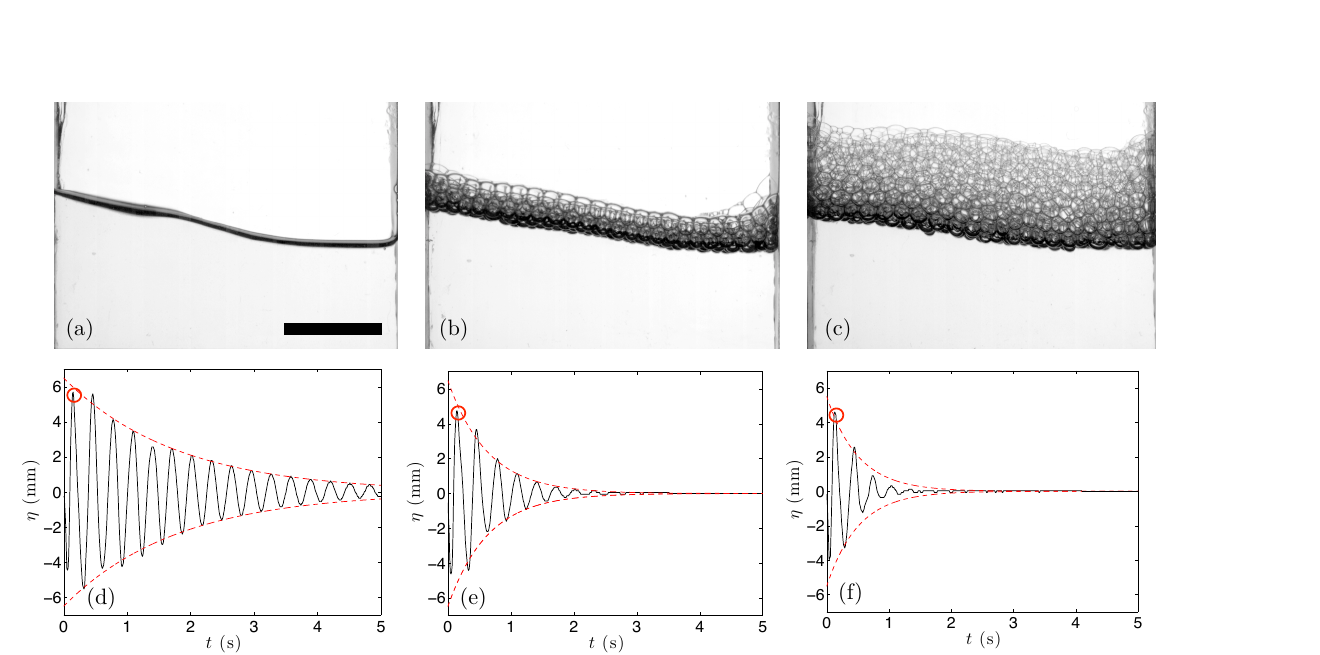}
           \caption{Oscillation of the interface for different foam thicknesses. (a-c) Visualization of the free interface after the impulse motion and (d-e) time-evolution of the free surface recorded 5 mm from the left vertical wall of the cell. The foam thickness varies: (a,d) no foam, (b,e) 6 mm of foam and (c,f) 20 mm of foam. The typical bubble diameter is $3\,{\rm mm}$. The red dotted lines in (d-f) show the exponential decrease of the amplitude of the generated wave. The red circles indicate the time at which the visualization (a-c) are made. Scale bar is $20\,{\rm mm}$.}
           \label{fig:2}
   \end{center}
\end{figure}

In Figure \ref{fig:2}(a-c), we present visualization of the sloshing of the free interface after the first oscillations following the impulse to the container. The liquid height is $H_1=40\,{\rm mm}$ and we use three thicknesses of foam, $H_0$. These observations and experiments performed with a harmonic forcing suggest that the maximum amplitude of the wave decreases with increasing the foam thickness. In addition, the time-evolution of the position of the interface is reported in Figure \ref{fig:2}(d-e). For the three foam thicknesses, the free-surface elevation follows a damped harmonic response $\eta(t)=\eta_0\,\exp(-t/\tau)\,\cos(2\,\pi\,f\,t)$. The frequency of the oscillation is independent of the foam thickness: $f=3.22\,\pm0.11\,\rm{s^{-1}}$ but the timescale $\tau$ over which the oscillations decay decreases with the foam thickness: $\tau=1.81\pm0.05\,{\rm s}$ without foam, $\tau=0.63\pm0.04\,{\rm s}$ for $6$ mm of foam and $\tau=0.37\pm0.01\,{\rm s}$ for $20$ mm of foam.

The increase of damping, $1/\tau$ is likely due to viscous dissipation as the foam is displaced along the sidewalls of the container. A systematic characterization of the influence of the foam and the different parameters of the system will be the subject of a forthcoming study.

Our work suggests the use of foam on top of a liquid to damp sloshing. The study offers potential applications in industrial processes such as the transport of liquefied natural gas in cargoes or the stabilization of propellant in rocket engines.  The proposed method requires stabilization of the foam which is traditionally achieved chemically \cite{stablefoam} or mechanically \cite{stablefoam2}.

\paragraph{Acknowledgements}
We thank Isabelle Cantat and Marie-Caroline Jullien for helpful discussions. This research was made possible in part by the CMEDS grant from BP/The Gulf of Mexico Research Initiative.

    \end{document}